\documentclass[a4paper,twocolumn,10pt]{article}
\usepackage{graphicx}
\twocolumn
\begin{document}
\pagestyle{empty}

\huge 
\title{ {\Huge Parrondo Paradox in Scale Free Networks} }

\tiny 
\author{   Norihito Toyota   \\
{\normalsize Faculty of Business Administration and Information Science } \\
 {\normalsize Hokkaido Information University, Ebetsu,069-8585, Japan} \\
{\normalsize Tel: +81-11-385-4411, Fax: +81-11-384-0134}\\
{\normalsize E-mail:toyota@do-johodai.ac.jp}  }  
\date{}
\maketitle
\normalsize
\thispagestyle{empty}
\bf
{\it Abstract--}
\small 
Parrondo's paradox occurs in sequences of games in which a winning expectation may be obtained by playing the games in a random order, 
even though each game in the sequence may be lost when played individually. 
Several variations of Parrondo's games with paradoxical property have been introduced.
In this paper, I examine whether Parrondo's paradox occurs or not in scale free networks. 
Two models are discussed by some theoretical analyses and computer simulations. 
As a result, I prove that Parrondo's paradox occurs only in the second model. 
\rm
\normalsize 

\begin{center}
I. \vspace{2mm} NTRODUCTION
\end{center}

\hspace{2mm} 
The original Parrondo's game consists of two losing games A and B where each is played by only one player.  
In the game A, only one biased coin is used, while in the game B two biased coins are used 
corresponding to the player's current capital.  
When a player plays individually each game, he/she  loses his/her capital on average. 
However, when the player plays two games in any combination, he/she always wins on average \cite{Harm1},\cite{Harm2}.    
Several variations of Parrondo's games apparently with such paradoxical property have been introduced;
 history dependence, one dimensional line, two dimensional lattice and so on\cite{Parr2}.\cite{Toral1},\cite{Miha1},\cite{Miha2}.
 
In this paper, I examine whether Parrondo's paradox occurs or not in scale free networks(SF networks)\cite{Albe2}, instead of two dimensional lattice.
This  is interesting as an empirical study, since scale free networks are a common  occurrence in our real world\cite{newBook}. 
Here two models are discussed by both some theoretical analyses and computer simulations. 
As a result, I prove that Parrondo's paradox occurs only in the second model. 
This paradox occurs by the quite different mechanism  from the original Parrondo's paradox.  
\begin{center}
II. \vspace{2mm} PARROND GAME ON SF NETWORKS
\end{center}
\hspace{2mm}  
Parrondo's paradox occurs in sequences of games in which a winning expectation may be obtained by playing the games in a random order, 
even though each game in the sequence may be lost when played individually.  
The original version of Parrondo's game consists of  the following two games and the initial capital of a player is $C(0)=0$; 

1. Game A: the probability  of winning is $P_A$ in this 

\hspace{2mm}   game. Usually $P_A<0.5$ is taken for losing game.   

2. Game B: If the capital $C(t)$ of the player at $t$ is 

\hspace{2mm}   a multiple of 3, the probability of winning is $P_B^{(1)}$, 
        
\hspace{2mm}   otherwise, the probability of winning is $P_B^{(2)}$.
        
3. Game A+B: Two games are mixed. The game A is 

\hspace{2mm}   played with probability $P$ and game B is played with 
 
 \hspace{2mm}  $1-P$. \\
In all there are 4 parameters,  $P$, $P_A$, $P_B^{(1)}$ and $P_B^{(2)}$,  controllable by  a planner of the game.  
When we win a game A or B or A+B,  we get one unit of capital and when we lose the game, we lose one unit of capital.   


 
The degree distribution of scale free networks is given by 
\begin{equation}
P(k)= \frac{(\alpha -1)}{ k_{min}^{1-\alpha} - k_{max}^{1-\alpha} }\;\; k^{-\alpha} \equiv C k^{-\alpha},  
\end{equation}
where $\alpha>0$ is an exponent in a power low, $C$ is a normalization constant and,  
$k_{min} $ and $k_{max} $ are the maximum degree and the minimum one in a network \cite{newBook}.  . 
In this paper, I construct scale free networks by using the preferential attachment (BA model) introduced by 
Barabashi et al.\cite{Albe2} with  $\alpha=3$ and $k_{min}=4$.      

When Parrondo's game is straightforwardly extended to the game in SF networks with degree $k$  according to \cite{Miha}, 
the number of parameters becomes $k+1$ and that is so large. 
Furthermore the degree differs in every node in scale free networks.
Then the game has too many parameters to analyze the game theoretically. 
So we introduce a cutoff $R$ or $r$ to introduce two types of models as the Parrondo game on SF networks by replacing the game B as follows. 

Model I: When there are not less than $R$ winners in players adjacent to a target player, 
the target player plays the game W whose winning probability is $P_W$, 
 and the target player plays the game L whose winning probability is $P_L $ in other cases  in the game B. 

Model II: When there are not less than $rk$ winners in players adjacent to a target player, 
the target player plays the game W whose winning probability is $P_W$, 
 and the target player plays the game L whose winning probability is $P_L $ in other cases  in the game B. 

Since Parrondo like paradox may accidentally occur in the capital of individual because of probability game,  
I analyze the average capital over all players as the time series of capitals in similar manner to the game on two dimensional lattice\cite{Miha}.


 \begin{figure}[t]
\includegraphics[scale=0.77,clip]{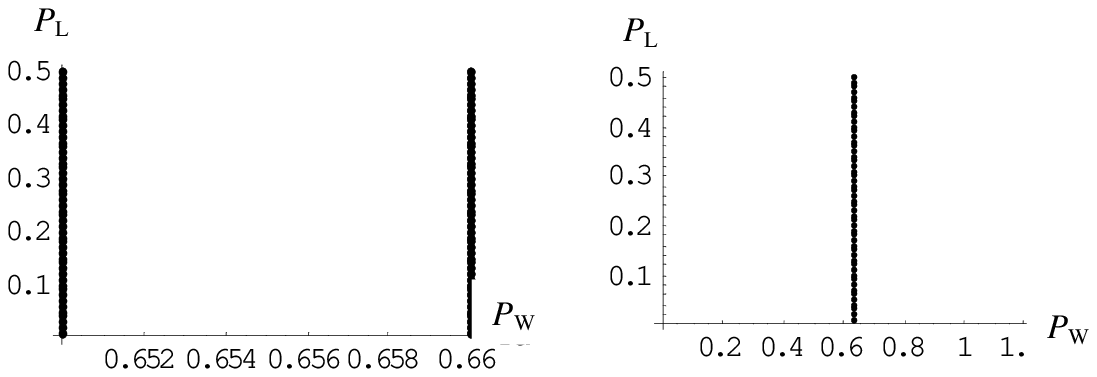} 
\small{{\small{Fig.1.  ($P_W$，$P_L$)  satisfying the conditions (5) and (6)  for $k_{max}=25$(left)  and for $k_{max}=65$(right) at $R=6$.   }}}
\vspace{4mm}
\begin{center}
\includegraphics[scale=0.77,clip]{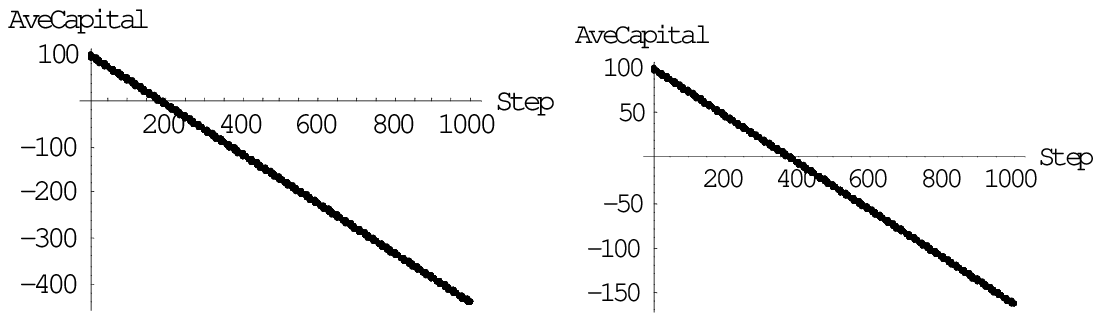} 
\end{center}
\small{{\small{Fig.2. The average capitals at  $R=5$，$(P_L, P_R)=(0.2,0.6)$  in the Game B (left) and the Game A+B (right) where $C0)=100$.   }}}
\end{figure} 


\normalsize

\begin{center}
{\it A.\vspace{2mm} Model I}
\end{center}
\hspace{5mm} 
I first put to the test by a computer simulation at the network size $N=400$, $P=0.5$ and $R=5$ in order to discern the outline of wide parameter space. 
A necessary condition for the paradox is 
$
(P_L-0.5)(P_W-0.5)<0. 
$
Computer simulations are made  where each player on the network asynchronously plays.     
As  results of the simulations, I found Parrondo's paradox does not occur in this parameter region.   
Furthermore, I found that the variation of $P_L$ and $P_W$ is only relevant to the variation 
of the absolute values of $C_{A+B}(t)$ but does not occasion any qualitative changes such as a reversal in capital. 
Much circumstantial evidence shows that the variation of $R$ is so. 
Thus the value of $R$ is not crucial.  


Next I introduce the discriminant for the node with degree $k$;
\begin{eqnarray}
D_{(k,P_{W,G} ,P_{L,G}  )} &=& \frac{ (1-P_{W,G})^{k-R} (1-P_{L,G} )^R }{ P_{W,G}^{k-R} P_{L,G}^{R} },   
\end{eqnarray}
where G takes A, B or A+B \cite{ToyotaPa}.
Notice that when the degree of  nodes  is smaller than the cutoff $R$, the players on the nodes necessarily play 
the game L in the game B. 
So for that case, the discriminant is 
\begin{eqnarray}
D_{(k,P_{L,B} )} &=& (1-P_{L,B} )^k/ P_{L,B}^{k}\;\;\;\mbox{for }k<R . 
\end{eqnarray}
The payoff of Game A+B is given as
 \begin{equation}
 P_{W(L),A+B}=PP_A+(1-P)P_{W(L),B}. 
\end{equation}

Players on nodes with high degree can play either the game W or the game L depending on the number of winners 
in neighboring players.
Taking account of the minimal degree $k_{min} =4$ in BA model adopted in this paper, 
I find a set of  conditions for paradox mean field approximation;  
\begin{eqnarray}
 C \Bigl( \sum_{k=4}^R  \; \frac{D_{(k,P_{L,B} )}}{k^{\alpha}} +  \sum_{k=R+1}^k  \; \frac{ D_{(k,P_{W,B} ,P_{L,B}  )} }{k^{\alpha}}  \Bigl) 
>1, \\
 C \Bigl( \sum_{k=4}^R  \;  \frac{D_{(k,P_{L,A+B} )}}{ k^{\alpha}  }+ \sum_{k=R+1}^k  \;  \frac{ D_{(k,P_{W,A+B} ,P_{L,A+B}  )}}{ k^{\alpha} } \Bigl) 
<1. 
\end{eqnarray}

 \begin{figure}[t]
\includegraphics[scale=0.65,clip]{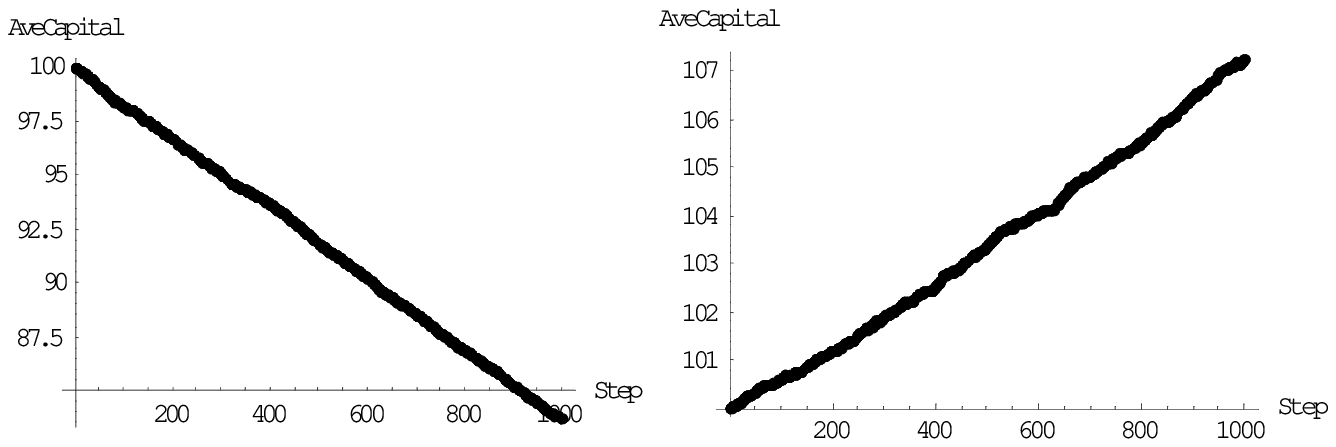} 
\small{{\small{Fig.3. Parrondo's paradox in Model II with ($r$,$P_W$，$P_L$)= (0.66,0.092,0.68),  $N=1000$ and $C(t=0)=100$.
The left is game B and the right is game A+B.  }}
}
\end{figure}

\normalsize
I found several parameter sets of ($P_L,P_W$) satisfying (5) and (6) by numerical calculations.   
A few examples of data points that satisfy  the conditions (5) and (6)  
in the parameter space $P_W$-$P_L$  are shown in Fig.1. 
I, however, could not confirm that paradox really does occur by numerical calculations in any parameters of them. 
A example of this is given in Fig.2 where Parrondo Paradox does not surely occur.   

\begin{center}
{\it A. \vspace{2mm} Model II}
\end{center}
\hspace{2mm}
In Model II, there is not necessity to introduce (3).    
Whether the Game W or L in the Game (A+)B is played depends on the winner's ratio adjacent to the target player. 
The conditions for Parrondo paradox in his game are given by 

\begin{eqnarray}
D_{(rk,P_{W,B} ,P_{L,B}  )} &>& 1,\\
D_{(rk,P_{W,A+B} ,P_{L,A+B}  )} &<&1.
\end{eqnarray}
In the parameter region that  $P_{W,B} < 0.5 <P_{L,B}$ and  $P_{W,A+B} < 0.5 <P_{L,A+B}$, they are
\begin{eqnarray}
P_{W,B} \Bigl(1+\frac{P_{L,B}}{ 1-P_{L,B} } \Bigr)^{1/\gamma} &<& 1,\\
P_{W,A+B} \Bigl(1+\frac{P_{L,A+B}}{ 1-P_{L,A+B} } \Bigr)^{1/\gamma}&>&1,
\end{eqnarray}
where $\gamma=(1-r)/r$. 
For the cases that $P_{W,B} > 0.5 >P_{L,B}$ or  $P_{W,A+B}  > 0.5 >P_{L,A+B}$, 
we have only to reverse L and W  in (9) and (10),  
since there is a sort of  duality symmetry: $L\leftrightarrow W$ and $\gamma \leftrightarrow 1/\gamma$ 
in the Game B.

So the necessary conditions that Parrondo's paradox occurs consist of  4 sets of conditions. 
One of necessary condition is
\begin{eqnarray}
1&>&P_{W,B} \Bigl(1+\frac{P_{L,B}}{ 1-P_{L,B} } \Bigr)^{1/\gamma}, \nonumber\\
P_{W,B} &<&0.5< P_{L,B} ,\;\;\;\;\;P+P_{L,B} >1,
\end{eqnarray}
Others are made by carrying out the duality transformation  $L\leftrightarrow W$ and $\gamma \leftrightarrow 1/\gamma$
for each of (11). 
Exploring parameters satisfying (11), we obtain  a parameter set 
$$ (P_A, P_{W,B},P_{W,L},r,P)= (0.48,0.092,0.68,2/3,0.50).$$  
They are candidate parameters for a paradox. 
Fig.3 actually shows that a paradox in the average capital occurs at the parameter set 
where $C(0)=0$ and the network size is $N=1000$.

However, since the conditions for the Parrondo's paradox have not any $k$ dependence from (9) and (10), 
  I notice that the wining probabilities for the Game B and the Game A+B are given by  
\begin{eqnarray}
P_B &=&r P_{W,B} +(1-r) P_{L,B}, \\
 P_{A+B} & =& pP_A+ (1-p) \bigr( rP_{W,B} +(1-r)P_{L,B} \bigl). 
\end{eqnarray}
Two kinds of probabilities for the occurrence of the paradox were needed for any nodes 
 in the original Parrondo game B and so the Game A+B. 
Now there is, however, no sign of the paradox in this case, because there is only one controlled wining probability of $P_B$.  
In fact substituting (12) into (13), we obtain
\begin{equation}
 P_{A+B} =P P_A+ (1-P) P_{B} = P(P_A-P_B)+P_B. 
\end{equation}
So $ P_{A+B} <0.5$ for $P=0.5$ or even $P<1$, $P_A<0.5$ and $0.5>P_B$. 
These consideration clearly show that both game B and game A+B are losing!.   

I investigate why the Parrondo's paradox occurred in the Fig.3.  
Notice that effective $r$ is always some rational number different from true $r$ for a special $k$, 
since $k$ takes only a natural number.
$r$ takes some effective value $r_{eff}$ due to the discrete nature of $k$.  
For example, when the possible winning ratios of a node with $k=4$ are practically $0/4$, $1/4$, $2/4$, $3/4$ and $4/4$. 
So choosing $r=2/3$,  $r$ becomes effectively $r_{eff}=r_{upper}=3/4$ for W game.    
 For L game, $r$ likewise becomes effectively $r_{eff}=r_{down}=2/4$. 
Thus two effective $r$s' appear!

For the same parameters as ones in Fig.3, the effective values of $r_{upper}$ and  $r_{down}$ can be estimated for each degree $k$. 
We can derive $P_{B,eff}$ and $P_{A+B,eff}$  from the formula substituted from $r$ to $r_{eff}$ in (12) and (13). 
The average values $<P_{B,eff}>$ and $<P_{A+B,eff}>$ are given by
\begin{eqnarray}
 <P_{B(A+B),eff}>&=& \int P(k) P_{B(A+B),eff}(k) \nonumber \\
    &=&\int C k^{-\alpha} P_{B(A+B),eff}(k).
 \end{eqnarray}
Since this value is sensitive to the normalization constant $C$, I refine parameters in the present model.   
I first use $\alpha =2.57 $ for the network really constructed in this paper, different from ideal value $\alpha=3$. 
Though $k$ can take infinite large value in itself, $k$ really takes from $k_{min}=4$, which is a restriction coming from BA model,
  to $k_{max}=30$ \cite{ToyotaPa}.  
  The values of $P_{B,eff}$  and $P_{A+B,eff}$ for each degree $k$ are described in Fig.4. 
 The average values of  $P_{B,eff}$  and $P_{A+B,eff}$  are $0.49$ and $0.503$, respectively,    
 when the normalization constant evaluated from this limited region of $k$ is corrected to $1.051$ times. 
 Then the capitals $C_B(t=1000)$ and $C_{A+B}(t=1000)$ are $90.8$ and $105.414$, respectively.  
 These values almost explain the capitals of the game B and the game A+B in Fig.3 well. 
 The exact correction for the normalization constant without any limit of $k$ is  $1.035$ which is comparable with the above value, 1.051. 
   
There are effectively multiple $r$ values, and so multiple $<P_{B,eff}>$ and $<P_{A+B,eff}>$,  
due to the various effective $r_{eff}$.  
Such the effect  creates Parrondo's paradox in the similar way as the original Parrondo's game,  
but not in different mechanism from the original one.   
Thus I call its paradox the second Parrondo's paradox.    

Thus I found that  Parrondo's paradox can occur in SF networks.  

 \begin{figure}[t]
\includegraphics[scale=0.85,clip]{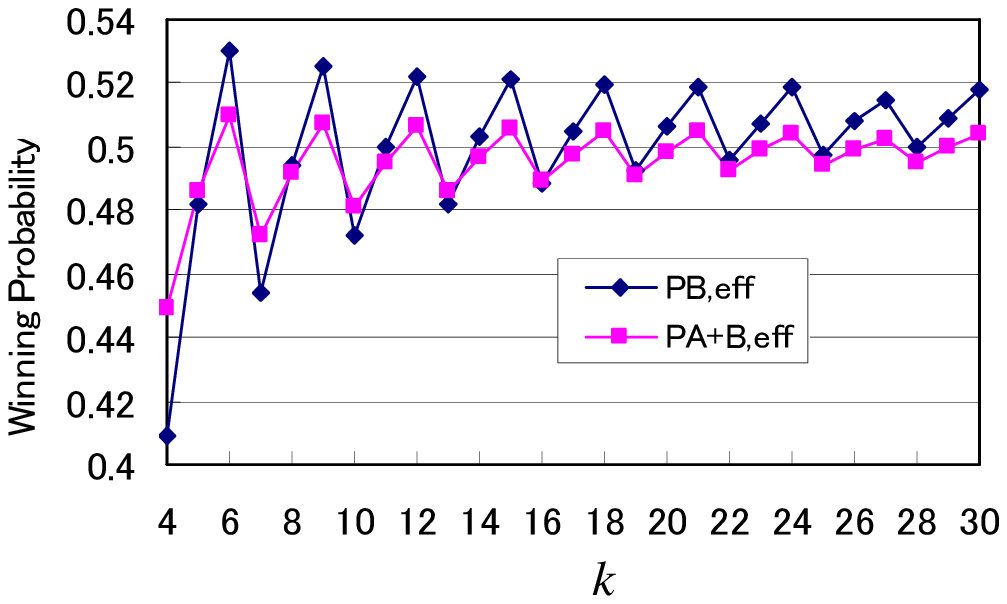} 
\vspace{4mm}
\small{{\small{Fig.4. Effective winning probabilities of the Game B and A+B in Model II with ($r$,$P_W$，$P_L$)= (2/3,0.092,0.68).  }}
}
\end{figure} 

\begin{center}
SUMMARY
\end{center}
\hspace{2mm}
In this paper, I explored whether paradox occurs or not in Parrondo's game on scale free networks which are more ubiquitous in real worlds than regular networks.   
It is too complicate to analyze the game in the general fashion, especially in giving theoretical considerations.  
So I consider only the case with the same number of parameters as the original Parrondo's game based on modulo $M=3$ in the capital.   
In this paper, the parameter corresponding to $M$ in the original Parrondo's game  is the cutoff $R$.    
When the number of winners adjacent to a target player is not more than $R$,  
the player plays Game L with the winning probability $P_{L,B}$ in Game B of Parrondo's game. 
Otherwise the player plays Game W with the winning probability $P_{W,B}$.
Two types of models are considered in this article. 
One is the the threshold game where $R$ takes a constant value independent of degree $k$. 
Another is the threshold ratio game where $R$ depends on degree $k_i$ of node $i$ such as $R_i=rk_i$. 
$r$ means the ratio of winners to all players adjacent to the player on node $i$.   

For the threshold  game, I accumulated circumstantial evidence that paradox does not occur by some computer simulations.
Furthermore I almost practically showed  that Parrondo's paradox did not occur in this naive case 
from theoretical point with numerical experiments of view.  
It, however, remains to be studied whether some paradox actually does not occur in excessively large scale networks. 

I showed that a paradox occurred for some parameter set in the threshold ratio model. 
In this case, the discrete nature of $k$ effectively induces plural winning probabilities in game B as the original Parrondo's game. 
It is thought that the various $P_{B,eff}$ or $P_{A+B,eff}$ appearing in this model create the paradox. 
The plural winning probabilities in game B are not given artificially as  the setting of the game but they 
are effectively induced by the discrete nature of $k$. 
So I called the paradox the second Parrondo's paradox. 

I never showed that it was hard to create Parrondo's paradox in scale free networks, generally. 
Notice that  the networks that  I studied is only BA models and only studied excessively naive setting in game B.  
I only focus my attention  on the numbers or the ratio of winners, and not on degree or the number of losers
(notice that considering both the number of losers and winners turns  out to consider degree and the number of winners as well). 
If making efficient use of  these information, we would find more rich aspects of Parrondo's  paradox in various types of scale free networks. 


\small

\end{document}